\documentclass[journal]{IEEEtran}

\usepackage{lipsum}
\usepackage{multicol}
\usepackage{graphicx}
\usepackage{amsmath}
\usepackage{graphicx}
\usepackage{cite}
\usepackage[utf8]{inputenc}
\begin{document}
\title{BiPMAP: A Toolbox for Predicting Perceived Motion Artifacts on Modern Displays}
\author{Guanghan~Meng*, Dekel~Galor*, ~\IEEEmembership{}Laura~Waller,~\IEEEmembership{}
        and~Martin~S. Banks ~\IEEEmembership{}
\thanks{*Guanghan Meng and Dekel Galor contributed equally to the work. }%
\thanks{All authors are with University of California, Berkeley}%
\thanks{Guanghan~Meng:~~guanghan\_meng@berkeley.edu}%
\thanks{Dekel~Galor:~~galor@berkeley.edu}%
\thanks{Laura~Waller:~~waller@berkeley.edu}%
\thanks{Martin~S.~Banks:~~martybanks@berkeley.edu}%
}%



\maketitle

\begin{abstract}
Viewers of digital displays often experience motion artifacts (e.g., flicker, judder, edge banding, motion blur, color breakup, depth distortion) when presented with dynamic scenes. We developed an interactive software tool for display designers that predicts how a viewer perceives motion artifacts for a variety of situations. We call it the Binocular Perceived Motion Artifact Predictor (BiPMAP). The tool enables the user to specify numerous stimulus, display, and viewing parameters. It implements a model of human spatiotemporal contrast sensitivity in order to determine which artifacts will be seen by a viewer and which will not. The tool visualizes the perceptual effects of discrete space-time sampling on the display by presenting side by side the expected perception when the stimulus is continuous compared to when the same stimulus is presented with the spatial and temporal parameters of a prototype display.  
\end{abstract}


%

\section{Introduction}
\IEEEPARstart{D}isplays are an interface that bridges the two ends of an information delivery system: the electronic signal supplied to the display and the viewer receiving the information converted to light by the display. For effective communication and a realistic visual experience, the interface should be tailored to the capabilities of the human visual system. Therefore, display manufacturers have worked to improve the design of their products---desktop monitors, projectors, televisions, cinema, and virtual-reality (VR) and augmented reality (AR) headsets---with the goal of achieving a realistic dynamic experience. 

On a display the motion of an object is presented as a sequence of static views. A proper design would ensure that the viewer perceives smooth motion free of artifacts \cite{watson1986window, watson2013high}. Unfortunately, this goal is often not achieved. Rather, a variety of motion artifacts can occur including flicker, judder, edge banding, motion blur, color breakup, and depth distortion \cite{watson1986window,hoffman2011temporal,johnson2014motion,daly2015psychophysical}. To predict when artifacts will appear and how to minimize them, one needs to consider the properties of the input image sequence, the spatial and temporal parameters of the display, and the spatio-temporal visual sensitivity of the viewer as he/she is positioned to view the display. Our software tool aims to make accurate predictions of the above motion artifacts, given a set of user-defined parameters, in order to predict performance for modern display designs.

To model the visual processing of the input stimulus we employ the contrast sensitivity function (CSF). The CSF characterizes the visual system's response to luminance variation in space and time. The CSF plots the reciprocal of the just-visible contrast at various spatiotemporal frequencies. Said another way, it defines frequency bands as transmissive---i.e., visible to the visual system---by defining the boundary between stimuli that are visible to a viewer and those that are not visible. The visible domain has been called the window of visibility \cite{watson1986window, watson2013high}. The motion on a display creates an output spatio-temporal frequency spectrum after filtering by the CSF. As long as that output spectrum is identical to the output spectrum of continuous real-world motion, the displayed object will be free of motion artifacts\cite{watson1986window}. Luminance has a significant effect on the CSF, so a prediction tool should also incorporate that parameter. Our model includes the effects of spatial and temporal frequency as well as luminance while remaining computationally simple for practical use.

We developed, for the first time to our knowledge, a 'click-and run' toolbox to guide the design of modern displays. We call it the Binocular Perceived Motion Artifact Predictor (BiPMAP). It enables prediction and visualization of motion artifacts. In this version of the toolbox, the input stimulus is a bright object moving horizontally at constant speed across a dark background \cite{watson1986window, hoffman2011temporal}. Our main contributions are: \par
\begin{enumerate}
\item A pipeline that visualizes a variety of predicted motion artifacts through analysis in the Fourier domain, making use of the window-of-visibility concept.
\item A simple CSF model incorporating spatial frequency, temporal frequency, and luminance.
\item Modeling smooth eye movements to determine their predicted effect on artifacts.
\item Predicting color breakup in color-sequential displays. 
\item Modeling binocular disparity in field-sequential stereo displays, enabling prediction of distortions in perceived depth.
\item An interface that accepts a comprehensive list of user-defined stimulus, display, and viewing parameters.
\item Toolbox as an interactive executable file with useful visualizations.

\end{enumerate}



\begin{figure*}[!t]
\centering
\includegraphics[width=6.5in]{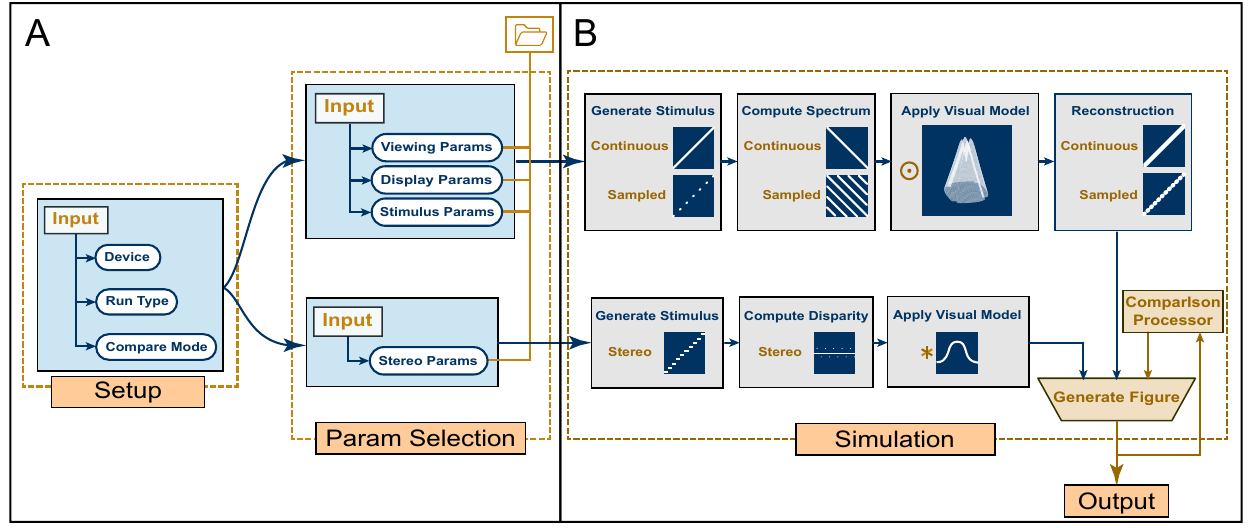}
\caption{Overview of BiPMAP. (\textbf{A}) Front end with user-defined parameters divided into two components: Setup (configuration inputs) and Parameter Selection (simulation variables). (\textbf{B}) Back-end pipeline for artifact predictions containing two streamlines---Non Stereo (top) and Stereo (bottom)---and a figure-generation step. When "Compare Mode" in the Setup component of (\textbf{A}) is enabled, the Comparison Processor integrates information from past runs and outputs a comparison figure.}
\label{flowchart}
\end{figure*}

\begin{figure*}[!ht]
\centering
\includegraphics[width=6.5in]{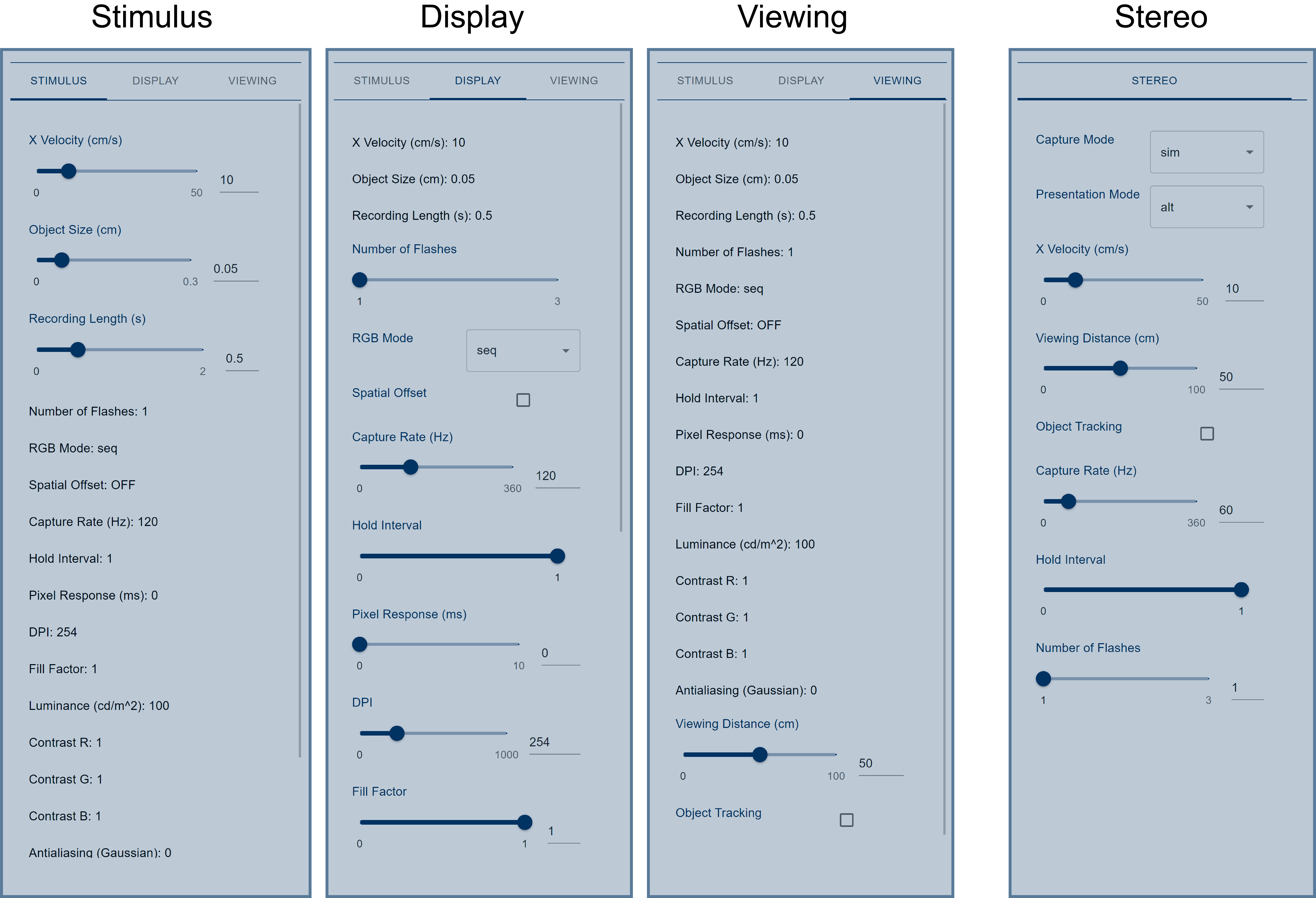}
\caption{Input parameters in the user interface. The three panels on the left show the interfaces for selecting, respectively, stimulus, display, and viewing parameters. Within each panel, only parameters of the specified group can be defined. The panel on the right shows the interface for stereoscopic displays.}
\label{ui}
\end{figure*}

\section{Background and related work}
\subsection{Window of visibility for the perception of a time-sampled moving stimulus}
A displayed moving stimulus is a time-sampled version of the corresponding real continuous motion. A useful method for representing and analyzing sampled motion was developed by Watson and colleagues \cite{watson1986window}. Here, we review this development, which forms the framework for our toolbox. 

For simplicity, consider an infinitely long vertical line with unit contrast moving horizontally at constant speed. The resulting contrast distribution along the vertical axis is uniform, while along the horizontal axis $x$ is: 
\begin{equation}\label{l_c}
    l(x,t)= \delta(x-rt),
\end{equation}
where $l(x,t)$ specifies the contrast as a function of position $x$ and time $t$. $r$ is the speed ($\Delta x$/$\Delta t$) and $\delta$ is the Dirac delta function. If the stimulus is presented stroboscopically (i.e., pixels are illuminated for a very short portion of each frame) on the display, the sampled stimulus $l_s(x,t)$ can be represented by the multiplication of the continuous contrast distribution and a sampling function:
\begin{equation} \label{l_s}
    l_s(x, t) = \Delta t\delta (x-rt)  \sum_{n=-\infty}^{\infty} \delta (t - n\Delta t),
\end{equation}
where $n$ is the frame number and $\Delta t$ is the sampling period. 

If the stimulus is illuminated for the full duration of each frame (sample and hold), the sampled stimulus now has a temporal staircase presentation: 
\begin{equation} \label{l_z}
    l_z(x,t) = \Delta t\delta (x-rt)  \sum_{n=-\infty}^{\infty} \delta (t - n\Delta t) )\ast z(x,t),
\end{equation}
where $z(x,t)$ is the unit staircase function [$\omega_s\delta(x)\text{rect}(t\omega _s)$] and $\ast$ denotes convolution and $\omega _s = 1/\Delta t$ is the sampling frequency.  

Watson and colleagues approached the problem of predicting motion artifacts in the Fourier domain. The frequency spectrum of the continuous stimulus is: 
\begin{equation} \label{L_c}
    L(u,\omega) = FT_{x,t}[l(x,t)] = \delta (ru+\omega),
\end{equation}
where $u$ is spatial frequency, $\omega$ is temporal frequency, and $FT$ is Fourier transformation.  The above frequency spectrum is a line of infinite length and has a slope equal to the opposite reciprocal of the motion velocity in the space-time domain. 

The stroboscopic sampled stimulus has a spectrum of: 
\begin{equation} \label{L_s}
    L_s (u, \omega) = FT_{x,t}[l_s (x,t)] = \sum_{n = -\infty}^{\infty} \delta (ru + \omega - n\omega_s),
\end{equation}
which contains an infinite set of replicates of the spectrum of the continuous motion with a spacing of $\omega_s$ between intercepts on the temporal frequency axis. 

For the staircase representation of a sampled stimulus, the Fourier transform yields the same function as Eqn. \ref{L_s} multiplied by a temporal sinc function (\text{sin}$\pi\omega/\pi\omega$) whose effect is to modulate the amplitudes of the replicates:
\begin{equation} \label{L_z}
    L_z (u, \omega) = \left[ \sum_{n = -\infty}^{\infty} \delta (ru + \omega - n\omega_s) \right] \text{sinc}(\frac{\omega}{\omega_s}). 
\end{equation}

To examine how the above frequency spectra are altered by the visual system, we use the CSF to quantify the resolution limits of the visual system in space and time. Watson and colleagues \cite{watson1986window} approximated the boundary between visible and invisible frequencies as a rectangle in the spatial and temporal frequency domain, and called this the window of visibility \cite{watson1986window}. As we said earlier, this conceptualization is very useful because it proposes that a discrete stimulus on a display will appear the same as a continuous stimulus whenever the frequency spectra after filtering by the CSF are the same.  

\subsection{Motion artifacts}
We consider several motion artifacts: 

\begin{itemize}
    \item \textit{Flicker}: 
    Flicker is perceived temporal variation in brightness. In the frequency domain, flicker will be perceived when replicates encroach the window of visibility near a spatial frequency of zero: i.e., when the replicates intersect the temporal frequency axis inside the window of visibility and have non-zero amplitude at that intersection \cite{hoffman2011temporal}. 
    \item \textit{Judder}: 
    Judder is an artifact in which motion appears unsmooth. It occurs when one or more of the sampling-induced spectral replicates fall within the window of visibility at non-zero spatial frequency.
    \item \textit{Edge banding}: 
    Edge banding is an artifact in which more than one instance of a moving edge is perceived. It often occurs along with judder. The likelihood of edge banding is greater when displays employ multi-flash protocols: repeated presentations of each frame\cite{johnson2014motion}. 
    \item \textit{Motion blur}: 
    Motion blur is an artifact in which a sharp edge is perceived as blurred\cite{watson2013high,klompenhouwer200654}. It occurs when the viewer makes a smooth tracking eye movement to follow the moving stimulus and the display is not stroboscopic. Judder and edge banding are not usually seen when motion blur is apparent. 
    \item \textit{Color breakup}: 
    Color breakup, also known as the rainbow effect, is an artifact observed with displays that present colors sequentially\cite{mori1999mechanism}: the leading and trailing edges of a moving object with broad wavelength distribution are perceived as distinct color fringes. Although the artifact is more evident with tracking eye movements, it can also be seen when the eye does not move and the object moves past. Color breakup can be minimized or even eliminated by various methods\cite{chen2009field, lin2010color, qin2019review} including applying spatial offsets to the 2\textsuperscript{nd} and 3\textsuperscript{nd} colors being presented\cite{johnson2014visibility}. 
    \item \textit{Depth distortion on stereoscopic displays}: 
    The above-mentioned motion artifacts occur in stereoscopic and non-stereoscopic displays alike. In addition, stereoscopic displays are frequently prone to another artifact: depth distortion\cite{burr1979binoculardelay,hoffman2011temporal,johnson2015stereoscopic}. Stereoscopic displays often present the images to the two eyes in temporal alternation (e.g., the left-eye's image is presented at one time and the right-eye's image at another time in that frame).  This temporal offset causes an alteration in the brain's estimate of the binocular disparity of a moving object, which in turn causes the object to appear at an unintended depth \cite{morgan1979perception, burr1979binoculardelay, read2005stroboscopic,hoffman2011temporal}. 
\end{itemize}

\section{Overall Design of BiPMAP}
The BiPMAP toolbox has the following design goals. It should allow for a comprehensive list of user inputs, including the velocity of the moving stimulus, and nearly all display and viewing parameters. It should adopt a simple but accurate CSF model. The toolbox should present continuous and sampled stimuli side-by-side, so that motion artifacts caused by the discrete sampling that accompanies displays can be identified, investigated, and possibly eliminated by adjusting design parameters. Finally, it should allow visualization of a variety of motion artifacts in non-stereoscopic and stereoscopic displays.

BiPMAP consists of an interactive user interface and computational pipeline (Figs.~\ref{flowchart} \& \ref{ui}). The executable toolbox is available at: https://github.com/CIVO-BiPMAP/executable/releases.

\subsection{Inputs and running modes}
\begin{itemize}
    \item \textit{Device selection}: BiPMAP automatically detects the computational devices available to the user (e.g., CPU, GPU) and lists them under the 'Device Selection' button (Fig.~\ref{flowchart}A, Fig.~\ref{ui}), allowing the user to select one. 
    \item \textit{Run type}: BiPMAP has two running modes: motion-artifact predictions for non-stereoscopic displays (default) and depth distortions for stereoscopic displays. Users can switch between the two modes with a toggle.  
    \item \textit{Compare mode}:
    When users want to compare results from different runs, BiPMAP can run under 'compare mode' which allows users to choose a master run that will be set as the reference during the comparison. By default---i.e., without selection of a master run---the perceived continuous motion will be the reference. 
\end{itemize}

\subsection{Motion-artifact predictions on non-stereoscopic displays}
The pipeline consists of four steps to compute and visualize artifacts: 
\begin{enumerate}
    \item Configure the continuous and sampled motion pipelines with user-defined parameters. 
    \item Compute the Fourier transforms of the continuous and sampled stimuli.
    \item Apply spatiotemporal filtering due to the CSF---i.e., a non-binary window of visibility---to obtain the output spectra.
    \item Reconstruct the perceived stimuli using inverse Fourier transforms with continuous and sampled representations side-by-side so that any motion artifacts can be easily identified. 
\end{enumerate}
\subsection{Input parameters \& stimulus configuration}
\hfill\par There are three sets of input parameters: 
Stimulus properties, display parameters, and viewing parameters (Fig.~\ref{ui}).  
\par 
\textit{Stimulus parameters} (Fig.~\ref{ui}, Stimulus): 

\begin{itemize}
    \item \textit{Velocity}: Input is in cm/s and is converted to \textdegree/s within the pipeline given the user-defined viewing distance. Direction of motion is horizontal.
    
    \item \textit{Stimulus size}: Stimulus width in cm along the dimension orthogonal to the motion axis. Converted into number of pixels and degrees using pixel size and viewing distances defined previously. 
    
    \item \textit{Recording length}: Duration of the sampling in seconds. Longer recording length improves spectral resolution but increases computation time; it is also bounded by the memory of the processing device. Default recording length is 0.5s. Length can be adjusted according to memory capacity and computational power of the GPU/CPU. When the toolbox is in RGB mode and the computational device is a GPU with \textless $24$GB of dedicated memory, we recommend decreasing the recording length. 

    \item \textit{$L_{max}$}: Luminance of the stimulus. Under RGB mode, it is the summed luminance from all three colors, and each color is assigned with a luminance of $\frac{1}{3} L_{max}$. 
\end{itemize}

\par
\textit{Display parameters} (Fig.~\ref{ui}, Display): 
\par
\begin{itemize}
    \item \textit{Number of flashes}: (number of flashes): This parameter specifies the number of times each frame is displayed before updating to new image data.
    
    \item \textit{RGB mode}: Three options are available: black and white (BW), field sequential (RGB-seq), and simultaneous (RGB-simul). If the user is not concerned with color artifacts, black-and-white mode is recommended to minimize processing speed and memory consumption. 
    
    \item \textit{Capture rate}: Frames per second (Hz), which is the number of samples along the motion trajectory per second. It differs from presentation rate when the number of flashes is more than 1. 
    
    \item \textit{Hold interval}: The proportion of the frame during which pixels are illuminated; it ranges from 0--1 (0 for stroboscopic; 1 for sample and hold). 
    
    \item \textit{Pixel response}: The time required for a pixel to reach full intensity and to return to the intensity before the frame started. The hold interval above is defined as the start of the rising pixel intensity to the end of the falling intensity. The rising and falling are both modeled as linear for computational simplicity. 
    
    \item \textit{DPI}: Pixels per inch. Pixel size is $1/DPI$. 
    
    \item \textit{Fill factor}: Pixel fill factor is the spatial proportion of the pixel that is illuminated; values from 0--1. Default value is 1. 

    \item \textit{Contrast}: Weber contrast, defined as:  
    \begin{equation} \label{contrast}
    Contrast = \frac{L_{max} - L_{min}}{L_{min}}, 
    \end{equation}
    where $L_{min}$ is the luminance of the background. from Eqn (\ref{contrast}), background luminance is calculated as:
    \begin{equation} \label{backgroundL}
    L_{min} = \frac{L_{max}}{1+ Contrast}, 
    \end{equation}
    Under RGB mode, the contrasts for three colors are assigned separately, and $L_{max}$ is replaced with $\frac{1}{3}L_{max}$. 
\end{itemize}
\par
\textit{Viewing parameters} (Fig.~\ref{ui}, Viewing)): 
\begin{itemize}
    \item \textit{Viewing distance}: Distance from the display to the viewer's eyes in cm. 
    
    \item \textit{Object tracking}: Checkbox to determine if the viewer's eye will track the moving stimulus. If checked, $v_{eye} = v_{stimulus}$ so the retinal speed of the stimulus becomes $0$. 
\end{itemize}

\subsection{Mathematical representation of user-defined stimuli}

Eqns. \ref{l_s}--\ref{L_z} set up the basic framework for our analysis. We next describe how we incorporated several additional parameters. Specifically, we provide details on how we expanded Watson's development to enable a comprehensive list of user-defined parameters. If the reader wishes to skip the mathematical development, go to Section E.

\hfill \par We first consider the situation in which no eye movement occurs. Starting from Eqn. \ref{l_c}, the contrast distribution from a narrow moving object with unit contrast can be expressed as: 
\begin{equation} \label{cd_c}
    cd_c(x,t) = \delta(x - rt) \ast ker(x),
\end{equation}
where $cd_c$ denotes the contrast distribution that is continuous in time and $ker(x)$ is the kernel that describes the spatial footprint of the object along the $x$ dimension: 
\begin{equation} \label{kernel}
\begin{split}
        ker(x) & = \frac{1}{X} \text{rect}(\frac{x}{X})\left[ \sum_{k = -\infty}^{\infty}\delta (x - kp)  \ast \frac{1}{fp} \text{rect}(\frac{x}{fp}) \right]\\
        & = \frac{1}{Xfp} \text{rect}(\frac{x}{X})\sum_{k = -\infty}^{\infty} \text{rect}(\frac{x-kp}{fp}),
\end{split}
\end{equation}
\noindent where $X$ is the angular object width, $k$ is the pixel index, $p$ is angular pixel size, and $f$ is the pixel fill factor. 

Temporal sampling of Eqn. \ref{cd_c} generates a stroboscopic representation of the moving stimulus: 
\begin{equation} \label{cd_s}
    cd_s(x,t) = \Delta t  \delta(x-rt) \ast ker(x) \sum_{n = -\infty}^{\infty}\delta (t - n\Delta t).
\end{equation}
We can use a generalized staircase function to incorporate the hold interval $h$ and pixel response $\tau$: 
\begin{equation} \label{z_g}
    z_g(t) = \frac{1}{(h\Delta t - \tau)\tau} \left[ \text{rect}(\frac{t}{h\Delta t - \tau}) \ast \text{rect}(\frac{t}{\tau}) \right],
\end{equation}
where the hold interval ($h$) contains both the plateau and the linear rising and falling of the pixel intensity defined by convolution. Incorporating this into the stimulus representation: 
\begin{equation} \label{cd_z}
    cd_z(x,t) = cd_s(x,t) \ast z_g(t).
\end{equation}
\par Applying the Fourier transform to Eqn. \ref{cd_z}, the spectrum of the configured stimulus is: 
\begin{equation} \label{FT_simple}
\begin{split} 
    CD_z(u,\omega) & = FT_{x,t} \left[ cd_s(x,t) \right] FT_{x,t} \left[ z_g(t) \right] \\ 
                   & = CD_s(u,\omega) Z_g(\omega) \\
                   & = \left[ \delta(\omega + ru)K(u) \ast\sum_{n = -\infty}^{\infty}\delta(\omega - n\omega _s) \right] Z_g(\omega),
\end{split} 
\end{equation}
where $K(u)$ is the Fourier transform of $ker(x)$ (Eqn. \ref{kernel}): 

\begin{equation} \label{K}
\begin{split}
    K(u) & = \text{sinc}(Xu) \ast \left[\text{sinc}(fpu)\sum_{k = -\infty}^{\infty}\text{exp}(-j2\pi kpu) \right],
\end{split}
\end{equation}
$Z_g(\omega)$ is the Fourier transform of the generalized staircase function (Eqn. \ref{z_g}):
\begin{equation} \label{Z_g}
    Z_g(\omega)  = \text{sinc} \left[ (\frac{h}{\omega_s} - \tau)\omega \right] \text{sinc}(\tau\omega).
\end{equation}

Plug Eqns. \ref{K} and \ref{Z_g} into Eqn. \ref{FT_simple}: 

\begin{equation}
\begin{split}
   CD_z(u,\omega) & = \sum_{k = -\infty}^{\infty} \text{sinc}(Xu) \ast \left[ \text{sinc}(fpu)  \text{exp}(-j2\pi kpu)\right] \\
                   & \sum_{n = -\infty}^{\infty} \delta (\omega + ru - n\omega_s) \\
                   & \text{sinc}\left[(\frac{h}{\omega_s} - \tau)\omega \right]\text{sinc}(\tau \omega).
\end{split}
\end{equation}

Thus far all frames are treated as black and white. The case for $RGB$ frames is similar, but with a few modifications. Each frame has three channels, and each channel has its own contrast distribution function in the same format but with different shifts in space and time. For the contrast distribution for all channels, the fill factor $f$ in equations below is replaced with $f_{RGB} = f/3$. In the simultaneous presentation mode of RGB colors:
    \begin{equation}
        cd_{simul}(x, t, i)  = cd_z(x - \frac{ip}{3}, t),
    \end{equation}
\noindent where $i$ is the index of color channels ($i = 0, 1, 2$ for red, green, and blue channels, respectively). 
    Therefore, the frequency spectrum has an additional phase shift: 
    \begin{equation}
        CD_{simul}(u,\omega,i) = CD_{z}(u,\omega) \text{exp}(-j\frac{2\pi ip}{3}u).
    \end{equation}
    In the sequential $RGB$ mode, the frame period $\Delta t$ in the staircase function $z_g(x,t)$ is replaced with $\Delta t/3$, and sampling frequency $\omega_s$ in its Fourier transform $Z_g(u, \omega)$ is replaced with $3\omega_s$. There is an additional shift in time: 
    \begin{equation} 
        cd_{seq}(x, t, i) = cd_z(x - \frac{ip}{3}, t - \frac{i\Delta t}{3})
    \end{equation}
    The resulting frequency spectrum is: 
    \begin{equation}
        CD_{seq}(u,\omega,i) = CD_{z}(u,\omega) \text{exp} \left[ -j\frac{2\pi i}{3}(pu + \frac{\omega}{\omega_s}) \right].  \;\;\;\;\;\;\;
    \end{equation}

\hfill \par Next, we consider the situation in which the viewer makes a smooth eye movement to track the stimulus. When tracking occurs, all captured frames produce the same contrast distribution on the retina at the frame onset. However, when the hold interval $h$ is greater than 0, the eye's motion introduces opposite motion on the retina during a frame. Therefore: 
\begin{equation}
    cd_{eye} (x,t) = \left[ ker(x)\ast \Delta t \delta(x+rt)  z_g(t) \right] \ast \sum_{n = -\infty}^{\infty} \delta(t - n \Delta t).
\end{equation}
The trajectory is continuous during frame intervals given the smooth motion of the eye. The spectrum of the stimulus is: 
\begin{equation}
\begin{split}
    CD_{eye} (u, \omega) & = \left[ K(u) \delta(\omega - ru) \right] \ast Z_g(\omega) \sum_{n = -\infty}^{\infty} \delta(\omega - n\omega_s)  \\
                         & =  K(u) Z_g(\omega - ru) \sum_{n = -\infty}^{\infty} \delta(\omega - n\omega_s) \\
                         & = K(u) \sum_{n = -\infty}^{\infty} Z_g(n\omega_s - ru)
\end{split}
\end{equation}
\par Similarly, the representation of $RGB$ stimulus incorporates the additional shifts and a different pixel fill factor. For the contrast distribution for each channel, the fill factor $f$ is now $f_{RGB} = f/3$.  
    In the simultaneous presentation mode of $RGB$ colors:
    \begin{equation}
        cd_{eye simul}(x, t, i)  = cd_{eye}(x - \frac{ip}{3}, t)
    \end{equation}
    where $i$ is the index of color channels which gives rise to a phase shift in the spectrum: 
    \begin{equation}
        CD_{eye simul}(u,\omega,i) = CD_{eye}(u,\omega) \text{exp}(-j\frac{2\pi ip}{3}u).
    \end{equation}
    The sequential $RGB$ mode is more complicated with object tracking. The additional time shift is accompanied by spatial offsets for different colors, and the staircase widths along both temporal ($\Delta t$ in $z_g(x,t)$) and spatial axes are reduced by a factor of 3: 
    \begin{equation}
    \begin{split}
        cd_{eyeseq}(x, t, i) & = \left[ ker(x)\ast \Delta t \delta(x+rt ) \right] \\
                    & z_g(t)\frac{3}{rh\Delta t} \text{rect} \left[ \frac{3x - irh\Delta t}{rh\Delta t} \right] \\
                   & \ast \left[\sum_{n = -\infty}^{\infty} \delta(t - n \Delta t) \right] \ast \delta (x - \frac{ip}{3})
    \end{split}
    \end{equation}
    The resulting frequency spectrum is: 
    \begin{equation}
    \begin{split}
        CD_{eyeseq}(u,\omega,i) & = K(u)Z_g(\omega - ru) \\
                                & \ast \left[ \text{sinc}(\frac{rh}{3\omega_s}u) \text{exp}(-j\frac{2\pi rh}{3\omega_s}u)\right] \\
                                & \text{exp}(-j\frac{2\pi ip}{3}u) \sum_{n = -\infty}^{\infty} \delta(\omega - n \omega _s) 
    \end{split} 
    \end{equation} 
    where the $\omega_s$ in $Z_g(\omega)$ is substituted by $3\omega_s$.

\subsection{CSF model} 

\begin{figure}[ht]
\includegraphics[width=3.4in]{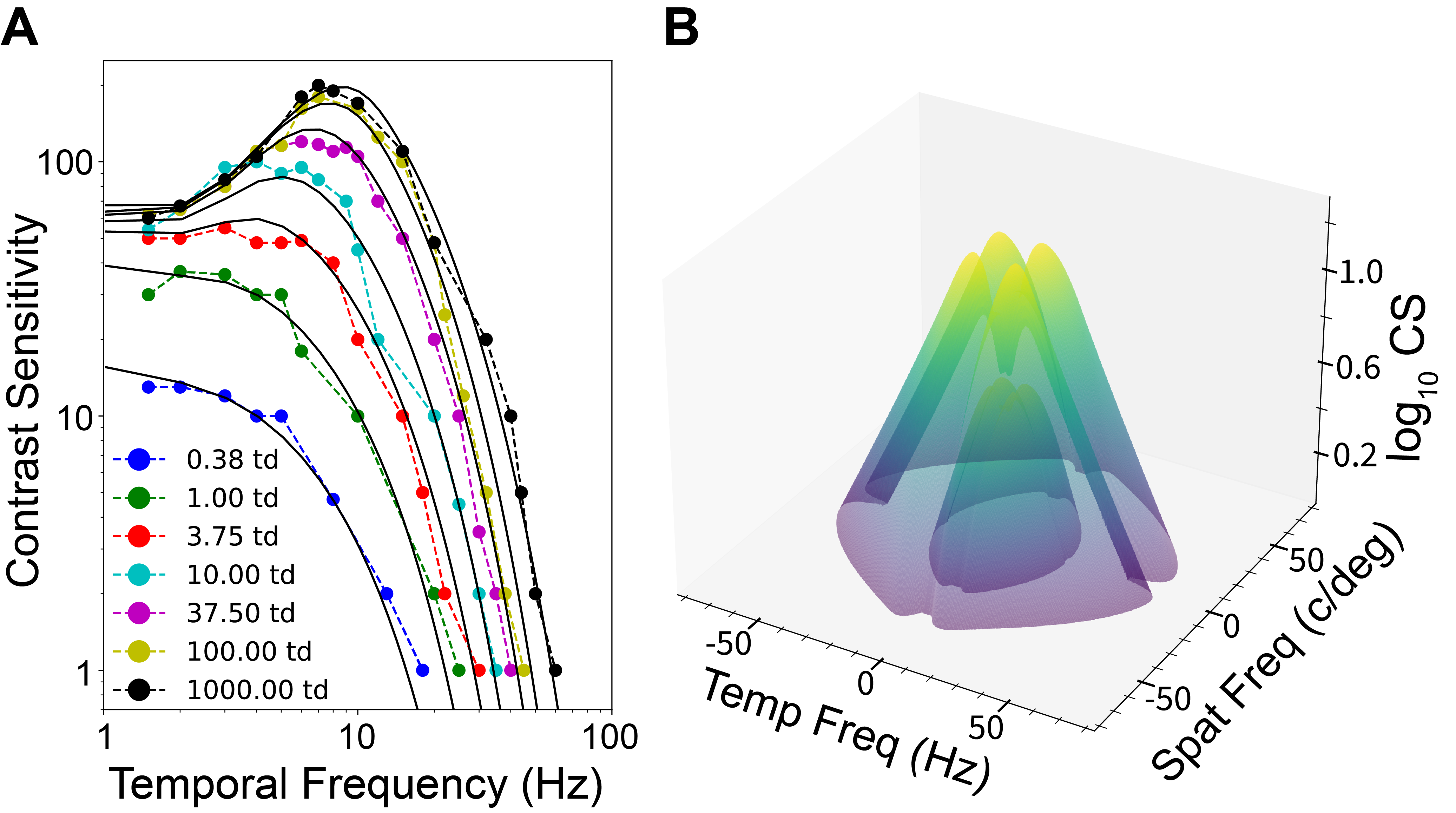}
\caption{Model of human contrast sensitivity function (CSF). (\textbf{A}) Empirical temporal CSF model (solid lines) obtained from fitting experimental data (round markers), plotted as a function of temporal frequency for different retinal illuminances (in trolands). (\textbf{B}) Dependence of spatiotemporal CSF on luminance. Log contrast sensitivity is plotted as a function of temporal ($\omega$) and spatial frequency ($u$) at two luminances: 0.5cd/m\textsuperscript{2} (inner cone) and 160 cd/m\textsuperscript{2} (outer cone).}
\label{csf}
\end{figure}

A binary rectangular window, i.e., the "window of visibility" proposed by Watson \cite{watson1986window}, is an oversimplification of the frequency response of the visual system. Other factors that affect sensitivity, such as luminance, should be considered as well. 
The "pyramid of visibility" is an upgraded model, also proposed by Watson and colleagues, in which high-frequency sensitivity was fit with linear models (in log space). The pyramid defines the CSF as a 3D surface with a height that increases with luminance (Fig.~\ref{csf}), and is only valid for high-frequency bands \cite{watson2016pyramid}. The stelaCSF model, in contrast, is more thorough but complicated with five dimensions \cite{mantiuk2022stelacsf}. In our CSF model, we incorporated three dimensions: spatial frequency, temporal frequency, and luminance. Previous work has shown that the CSF is approximately separable at high frequencies, so we assumed separability to simplify computation. Our CSF model is: \textbf{}
\begin{equation} \label{csfequation}
    CS(u, \omega) = \sqrt{CS_s (u) CS_t (\omega)},
\end{equation}
\noindent where $CS_s(u)$ is the spatial CSF and $CS_t(\omega)$ is the temporal CSF. 
The parameters of the spatial model were determined from previous work \cite{barten2003formula}: 
\begin{equation} \label{spatialCSF}
    CS_s(u) = \frac{5200e^{-0.0016u^2(1 + 100/L)^{0.08}}}{\sqrt{(1 + \frac{144}{X_o^2} + 0.64u^2)(\frac{63}{L^{0.83}} + \frac{1}{1-e^{-0.02u^2}} )}},
\end{equation}
where $X_o$ is the angular object size (deg), $L$ is luminance (cd/m$^2$), and $u$ is spatial frequency (cpd). 
To our knowledge, there is no temporal CSF model that incorporates luminance, so we developed one by fitting an empirical function to a set of published psychophysical data \cite{watson1986temporal, deLange1958} (Fig.~\ref{csf}A):

\begin{equation}\label{temporalCSF}
    CS_t(\omega) = \frac{5360L^{2.51}e^{-0.16\omega^{L^{-0.017}}}}{\sqrt{(\frac{a\omega^{bL}}{L^{-4.98}} + 1)(\frac{c}{L} + \frac{d}{1.007-e^{f\omega^{3.8}}})-L^5}},
\end{equation}
\noindent where $a = 2.1e+9$, $b = 9e-4$, $c = 1.2e-7$, and $d = -2.7e-4$. The original data were in terms of retinal illuminance (trolands) rather than luminance (cd/m$^2$), so we converted using an existing formula \cite{barten2003formula}.

From Eqns. \ref{csfequation}--\ref{temporalCSF}, we obtained our spatiotemporal CSF model. As expected, peak contrast sensitivity grows with increasing luminance. For example, it increases by a factor of 4 as luminance increases from $0.5$ to $160\mathrm{cd/m^2}$ (Fig.~\ref{csf}B). The specific CSF profile for the configured stimulus is determined based on the average luminance across the viewer's fovea: 
\begin{equation} \label{L_mean}
    L_{mean} = \frac{L_{max} X D_{fovea} + L_{min} (S_{fovea} - X D_{fovea})}{S_{fovea}}
\end{equation} 
where $D_{fovea} = 5 $\textdegree$ $ is the angular fovea diameter \cite{wandell2013}, and $S_{fovea}$ is the area of the fovea calculated as: 
\begin{equation}
    S_{fovea} = \frac{\pi D_{fovea}^2 }{4}
\end{equation} 

The predicted stimulus based on the input's Weber contrast is then determined using the CSF model mentioned above. 

\subsection{Depth distortion in stereoscopic displays}
In the stereoscopic mode, the user inputs parameters involved in calculating disparity. The right panel in Fig.~\ref{ui} shows the user inputs. 

\par
\textit{Parameters in binocular disparity calculation}:

\begin{itemize}
    \item  \textit{Capture mode}: Simultaneous or alternating left- and right-camera capture. Simultaneous capture is the default because it is more common. 
    
     \item \textit{Presentation mode}: Simultaneous or alternating left- and right-eye presentation. Alternating presentation is the protocol for temporal-interlaced stereoscopic displays, which are fairly common. Alternating is the default. 
\end{itemize}

\begin{figure*}[!t]
    \centering
    \includegraphics[width = 6.6in]{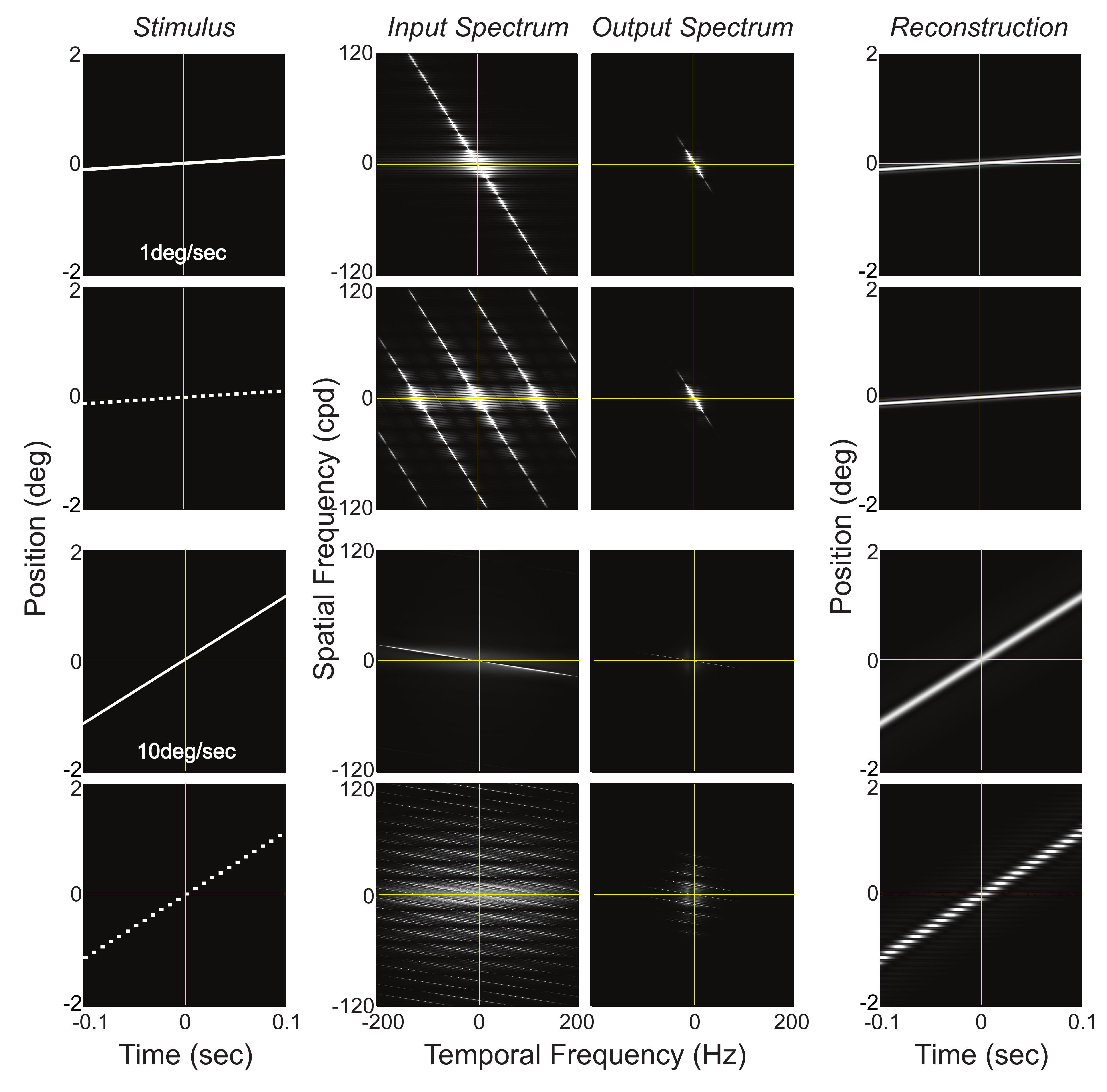}
    \caption{Effect of stimulus speed on judder. Stimulus column: The stimulus is a bright line whose position is plotted as a function of time. The user specifies line width and speed. Input spectrum column: 2D discrete Fourier transform of the stimulus. Spatial frequency and temporal frequency are plotted on the ordinate and abscissa, respectively. Brightness represents amplitude. Output spectrum column: Spectrum filtered by the CSF. Reconstruction column: Perceived stimulus reconstructed using inverse Fourier transform of the output spectrum. Top two rows: Results for continuous (first row) and sampled stimulus (second row) with a speed of 1cm/s (1.15\textdegree/s). No motion artifacts perceived from the sampled stimulus. Bottom two rows: Results for continuous (third row) and sampled stimulus (fourth row) with a speed of 10cm/s (11.42\textdegree/s). Pixel density: 300dpi. Capture rate: 120Hz. Hold interval: 0.5. Viewing distance: 50cm. }
    \label{velocity}
\end{figure*}

\begin{figure*}[!t]
    \centering
    \includegraphics[width = 6.6in]{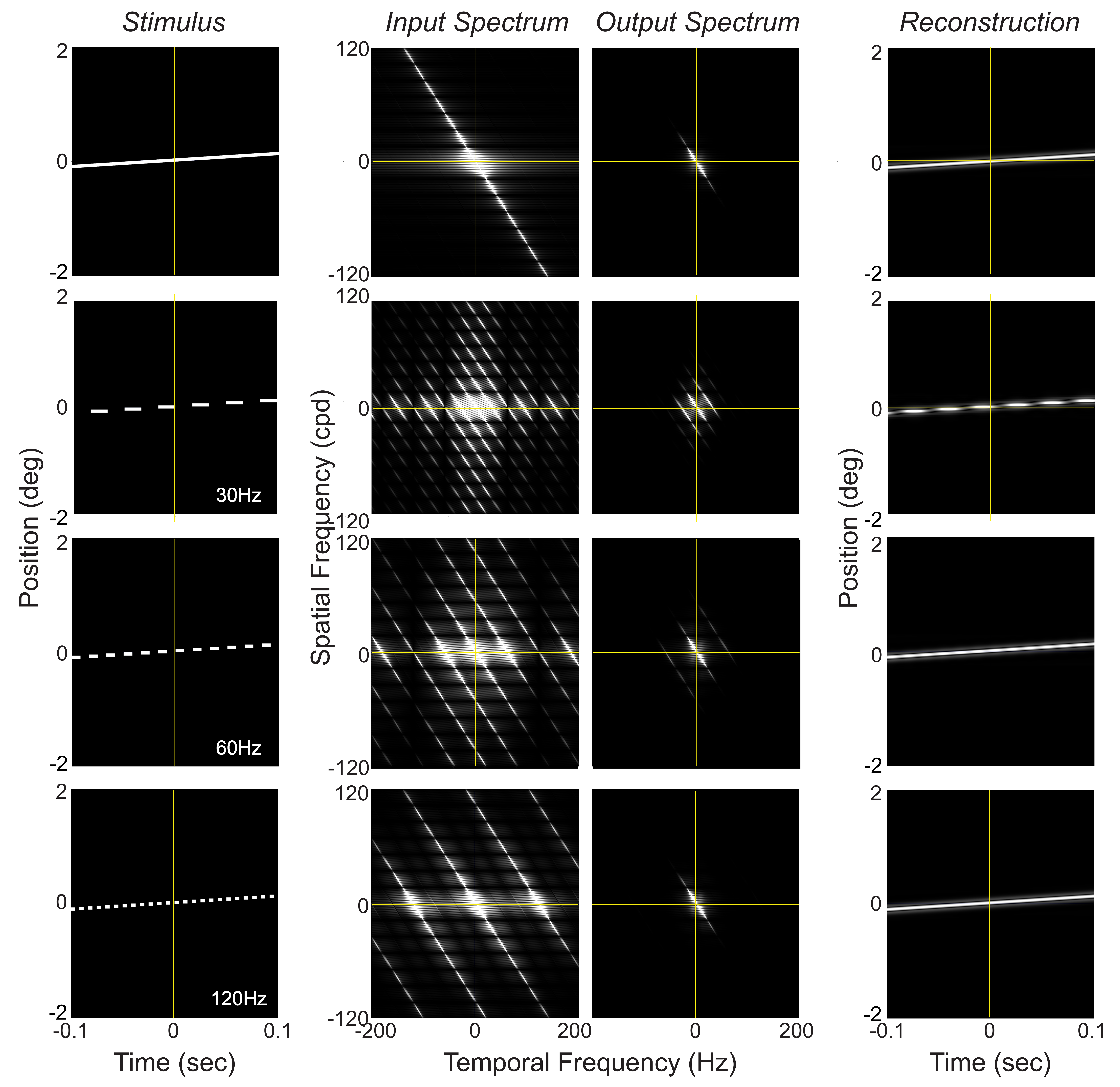}
    \caption{Effect of capture rate on judder. Columns in same format as Fig.\ref{velocity}. Top row: Results for continuous stimulus. Second row: Output from sampled stimulus with $30$Hz capture rate. Third row: Same but with capture rate of $60$Hz. Fourth row: Same but with capture rate of $120$Hz. Stimulus speed: 1cm/s (1.15\textdegree/s). Pixel density: 300dpi. Hold interval: 0.5. Viewing distance: 50cm.}
    \label{framerate}
\end{figure*}

\begin{figure*}[!t]
    \centering
    \includegraphics[width = 6.6in]{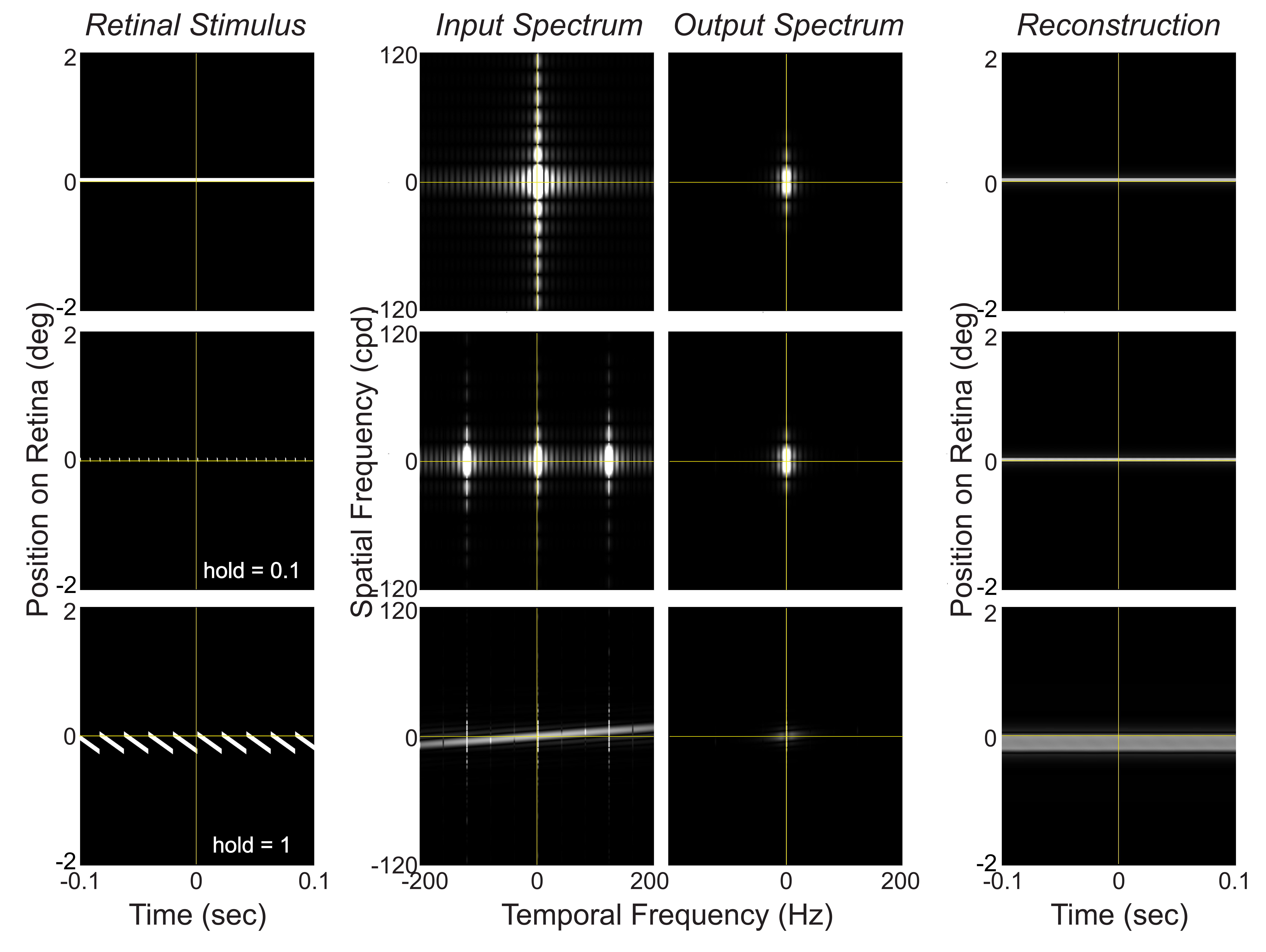}
    \caption{Effects of hold interval and tracking eye movement on motion blur. Columns in same format as Figures \ref{velocity} and \ref{framerate}. Stimulus position and spatial frequency are in retinal coordinates. Viewer makes a tracking eye movement to fixate the moving stimulus. The continuous stimulus (row 1) is therefore stationary on the retina. Hold interval in the sampled stimulus is 0.1 and 1 in the second and third rows, respectively. Motion blur occurs with large but not small hold interval. Stimulus velocity on display: 20cm/s (22.62\textdegree/s). Pixel density: 300dpi. Capture rate: 120Hz.Viewing distance: 50cm.}
    \label{motionblur}
\end{figure*}

\begin{figure*}[!t]
    \centering
    \includegraphics[width = 3.5in]{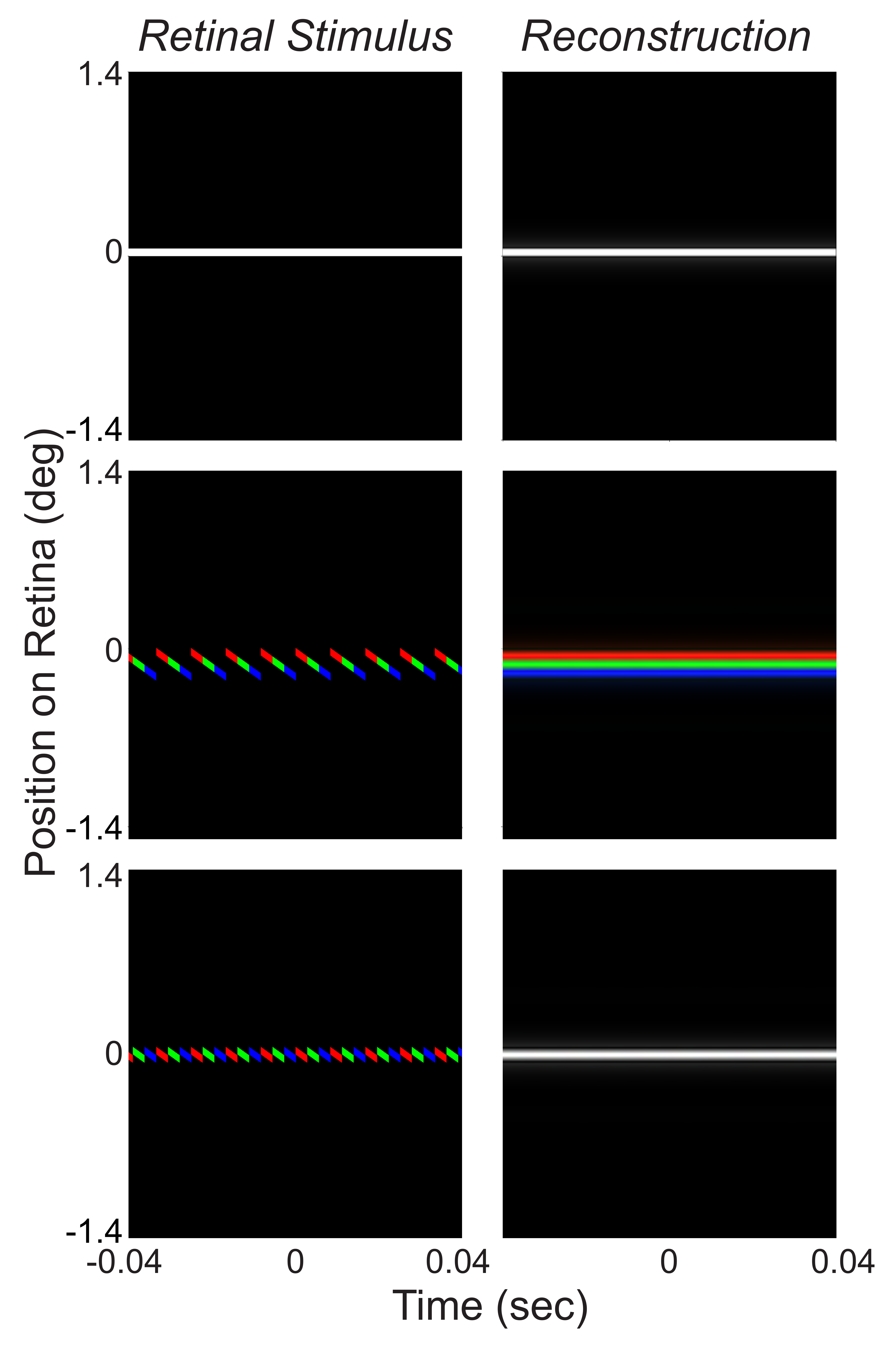}
    \caption{Color breakup with tracking eye movement. Left column: Stimulus moving at 20cm/sec (22.62\textdegree/s) with tracking eye movement. Position on the retina is plotted as a function of time. Right column: Reconstructed stimuli. Top row: Continuous stimulus with red, green, and blue presented simultaneously. Middle row: Sampled stimulus presented with color-sequential display. Bottom row: Sampled stimulus with spatial offsets to counter color breakup. Stimulus speed: 20cm/sec (22.62\textdegree/s). Pixel density: 300dpi. Capture rate: 120Hz. RGB mode: sequential. Hold interval: 1. Viewing distance: 50cm.}
    \label{colorbreakup}
\end{figure*}

\section{Task-based functionality of BiPMAP}
\subsection{Judder}
From Eqns \ref{l_c}--\ref{L_z}, increasing speed produces counterclockwise rotation in the space-time domain and horizontal shear in the frequency domain. Fig.~\ref{velocity} shows this for speeds increasing from $1$--$10$cm/s. Slope in space-time is greater for the faster speed (Fig.~\ref{velocity}, left column), and shallower in the Fourier domain (Fig.~\ref{velocity}, second column). The third and fourth columns show, respectively, the spectra after filtering by the CSF, and perceived stimuli after filtering. If we assume a capture rate of 120Hz, only one component of the frequency spectrum for the slower sampled stimulus is transmitted through the CSF (Fig.~\ref{velocity}, second row, filtered spectrum), and that component is the same as the one in the continuous stimulus. In other words, the output frequency spectra of the continuous and sampled stimuli are identical (Fig.~\ref{velocity}, first and second rows, filtered spectrum), which means that the sampled stimulus should be perceived as moving smoothly (Fig.~\ref{velocity}, first and second rows, Reconstruction). In contrast, the faster stimulus produces more than three replicates that fall within the window of visibility (third and fourth rows). As a result, the reconstructed stimulus has gaps, which means that the perceived motion will be discontinuous: i.e., judder will occur. Note that the filtered spectra for the slower and faster stimuli only intersect the temporal frequency axis at the origin, which means that there will be no visible flicker despite the presence of judder. 

The occurrence of judder and not flicker is also revealed in Fig.~\ref{framerate}. In this set of examples, the speed is 1cm/s while the capture rate increases from 30Hz (Fig.~\ref{framerate}, second row) to 120Hz (bottom row). Greater capture rates---i.e., sampling rates---push the replicates in the sampled frequency spectra farther from one another until only one remains inside the window of visibility. Thus at sufficiently high capture rates, the reconstructed stimulus (Fig.~\ref{framerate}, bottom row) should appear to move smoothly (top row). 

When the hold interval is increased from a small value ($\sim$0) to a large one ($\sim$1), the temporal sinc function attenuates higher frequencies in the replicates, but that attenuation occurs mostly outside the window of visibility. Thus increasing the hold interval does little to suppress judder. 

\subsection{Motion blur}
When viewers track a moving continuous stimulus, the velocity on the retina is zero, but because the brain measures the eye motion through extra-retinal signals, the stimulus will appear to be moving and doing so smoothly. When viewers track a sampled stimulus of the same velocity, they perceive the motion but also often experience motion blur \cite{klompenhouwer200654}. The hold interval is crucial here. When the interval is large, blur is experienced because the static image of the stimulus in one frame is smeared across the retina as the eye keeps moving \cite{feng2006lcd}. This is illustrated by Fig.~\ref{motionblur}, bottom row. Notice the diagonal lines in the left panel, which is the smearing across the retina during the hold interval. The limited spatiotemporal bandwidth of the visual system acts like a low-pass filter and causes a sharp contour to appear blurred. Reducing the hold interval causes less retinal smear and therefore motion blur is minimized or even eliminated (Fig.~\ref{motionblur}, second row). 
\begin{figure*}[!t]
    \centering
    \includegraphics[width = 6.5in]{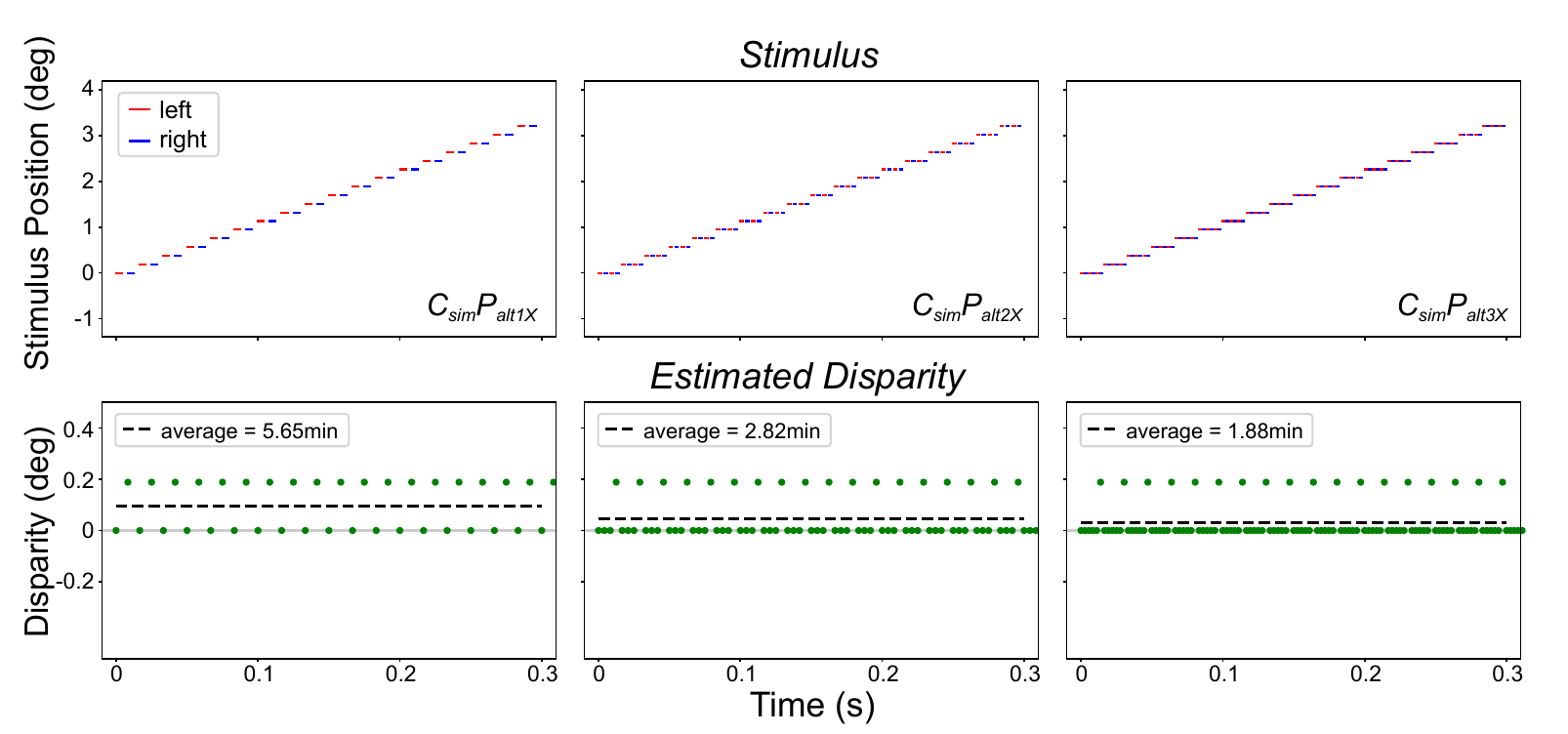}
    \caption{Estimated disparity with different display protocols. Labels for each column indicate that capture ($C$) is simultaneous and presentation ($P$) is sequential. The number followed by $X$ is the flash number. Top row: Simultaneous capture and sequential presentation with different flash numbers for a stimulus with a nominal disparity of zero moving left to right. Position on the display is plotted as a function of time. Red lines indicate the left-eye presentation and blue lines the right-eye presentation. The first, second, and third panels show the displayed stimulus for single-, double-, and triple-flash, respectively. Bottom row: Estimated disparity for these protocols. Estimated disparity is plotted as a function of time. Green dots represent the disparities associated with left- and right-eye pairings in which right eye leads or lags the left eye. Dashed line indicates the estimated disparity. Stimulus speed: $10$cm/s (11.31°/s). Capture rate: $60$Hz. Viewing distance: 50cm.}
    \label{disparity-flash}
\end{figure*}

\subsection{Color breakup in field-sequential RGB displays}
Color breakup is another artifact that affects the viewing experience. It occurs when the display presents colors sequentially as with most DLP projectors \cite{hornbeck1997digital}. Breakup is most noticeable when viewers track the displayed stimulus with smooth eye movements (Fig.~\ref{colorbreakup}, middle row), but it can also be seen when the eye is stationary and the stimulus moves past \cite{mori1999mechanism}. When red, green, and blue are presented sequentially within a frame and the viewer tracks the stimulus, each color will fall on a slightly different part of the retina. As a result, a white line on a black background will be imaged on the retina as three spatially separated lines: red, green, and blue; red will appear farther in the direction of motion than green, which will appear farther than blue. Thus, color breakup is an appearance of spatial color fringing caused by offsets in time. One can apply spatial offsets to eliminate the artifact \cite{johnson2014visibility}. The spatial offset required for each color channel is:
\begin{equation}
    \textit{offset}_i = (i - 1)\frac{v}{3c},
\end{equation}
\noindent where $i$ is the presentation order for the color channel that is being corrected, ranging from $1$--$3$, $v$ is stimulus speed, and $c$ is the capture rate. The effect of adding such offsets is shown in Fig.~\ref{colorbreakup}.
When we add spatial offsets to green and blue, color breakup is successfully eliminated (Fig.~\ref{colorbreakup}, bottom row).

\subsection{Depth distortion in stereoscopic displays}

When filming or generating content for stereoscopic display, the left and right images are usually captured simultaneously. Many stereoscopic displays then present the left and right images in temporally alternating fashion to the two eyes \cite{park2014effect}. As a consequence, horizontally moving objects can appear to be displaced in depth; this is a depth distortion \cite{kim2015overview}. To understand the cause of this effect, we must look into how the brain estimates binocular disparity. Consider a situation in which the disparity of a moving object is intended to be zero (meaning the left- and right-eye images are the same and the object is meant to appear at the depth of the display screen) and the left- and right-eye images are presented in alternation in a given frame. The brain must pair the left and right images in order to compute the binocular disparity. The problem for a given left-eye image is whether to pair it with the following right-eye image or the preceding one. If the left-eye image occurs in the first half of the frame, the pairing with the following image is correct and the pairing with the preceding one is incorrect. But the brain has no way to know which pairing is the correct one, so it matches using an average of the two \cite{read2005stroboscopic,banks20163d}. Specifically, the left-eye image is paired with both the preceding and following right-eye images and the disparity is derived from the average of those two pairings. The estimated disparity is therefore incorrect, which produces an apparent displacement of the object in depth. BiPMAP allows the user to determine what the estimated disparity is likely to be for different stimuli and display protocols. 

Binocular disparity is the position of a feature in the right eye relative to the same feature in the left eye. Specifically, 

\begin{equation}
d = x_r - x_l,
\label{disparity_definition}
\end{equation}

\noindent where $x_r$ and $x_l$ are the horizontal coordinates of the right- and left-eye images, respectively. When $d$ is positive, the disparity is uncrossed, which means that the feature should appear farther than the display screen.

The disparity error for a variety of stimulus speeds, capture rates, and flash numbers is: 

\begin{equation}
e = \left({\frac{v}{c}}\right)\left(\frac{1}{2f}\right),
\label{disparity_error}
\end{equation}

\noindent where $e$ is the error in degrees, $v$ is horizontal speed in \textdegree/s, $c$ is capture rate in Hz, and $f$ is flash number. When $e$ is positive, the object has unintended uncrossed disparity and should appear farther than desired. 

Consider an object moving from left to right with zero disparity (i.e., it should be seen in the plane of the display). When the left-eye image is presented before the right-eye image in each frame, the estimated disparity will be the average of 0 (pairing with the following image) and -${\Delta}x$ (pairing with preceding image). So instead of obtaining the correct estimate of zero disparity, the brain obtains an incorrect estimate of -${\Delta}x/2$, which in the example shown in the left column of Fig.~\ref{disparity-flash} produces an average disparity error of 5.65minarc, which is readily visible because it is more than an order of magnitude greater than the disparity threshold. The error can be derived from Eqn. \ref{disparity_error} by substituting $-{\Delta}x/{\Delta}t$ for $v$, $1/{\Delta}t$ for $c$, and $1$ for $f$: 
\begin{equation}
e = \left(\frac{-{\Delta}x}{{\Delta}t}\right)\left(\frac{{\Delta}t}{2}\right)
= -\frac{{\Delta}x}{2}.
\end{equation}
Because of this, the object's perceived distance is nearer than it should be: i.e., a depth distortion. If the motion was right to left, the object would appear farther than it should. 

This depth distortion can be eliminated, of course, by presenting the images to the two eyes simultaneously or by capturing the left- and right-eye image data sequentially at the same alternation rate as will be used in the presentation.

Field-sequential presentation is commonplace in stereoscopic cinema and is usually accompanied by multi-flash. For example, in the RealD protocol each pair of images is presented three times ('triple flash') before updating to a new pair of images \cite{cowan2008real}. This reduces the magnitude of the expected depth distortion by a factor of 3 as shown by Eqn. \ref{disparity_error} and the right panel of Fig.~\ref{disparity-flash}. The expected effect with double flash is smaller and illustrated by the middle panel of the figure. 

\begin{figure}[ht]
    \centering
    \includegraphics[width = 3.3in]{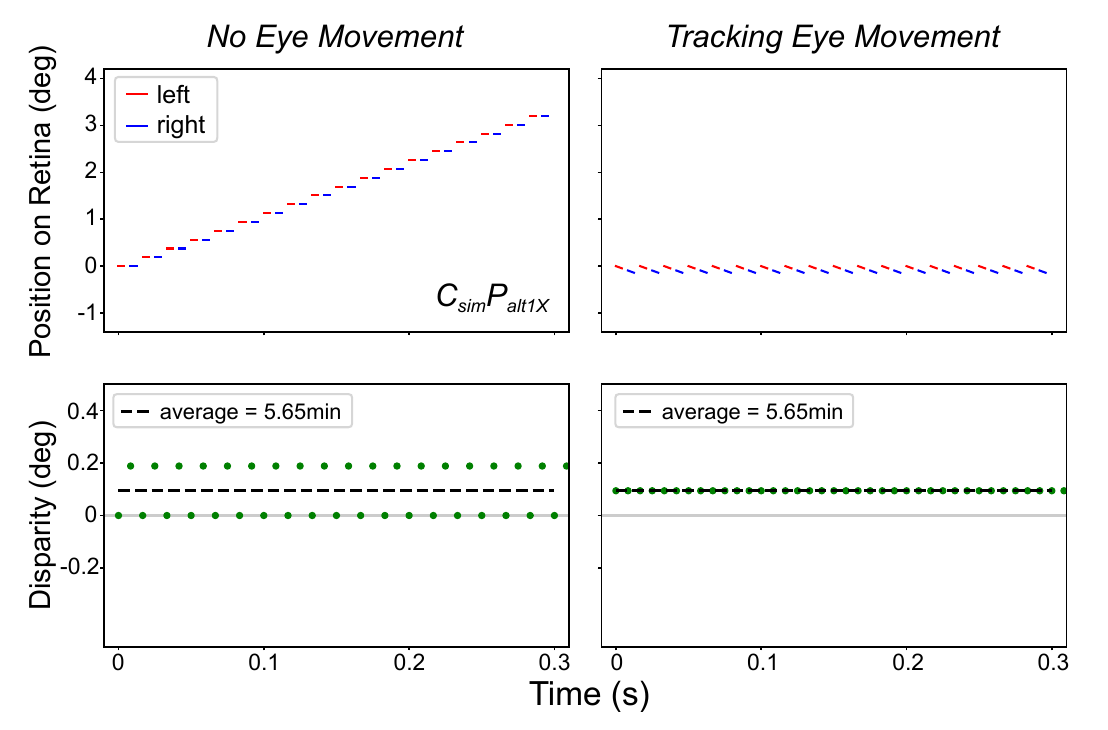}
    \caption{Effect of eye motion on estimated disparity. \textbf{(A)} Stimulus (top) and estimated disparity (bottom) with no tracking eye movement. \textbf{(B)} Stimulus (top) and estimated disparity (bottom) with tracking eye tracking. Top row: Stimulus on display with simultaneous capture and alternating presentation, single flash. Bottom row: Estimated disparity. Green dots represent the disparities associated with left- and right-eye pairings in which right eye leads or lags the left eye. Dashed line is the estimated disparity. The estimated disparity is the same with and without eye tracking. Stimulus speed: $10$cm/s (11.31°/s). Capture rate: $60$Hz. Viewing distance: 50cm.}
    \label{disparity-eyemotion}
\end{figure}

When the viewer tracks the horizontally moving object with a smooth eye movement, there is no effect on the depth distortion created by alternating presentation. This can be shown by running BiPMAP in stereo mode with eye tracking enabled (Fig.~\ref{disparity-eyemotion}).

\section{Conclusions and Future Direction}
We developed a toolbox---BiPMAP---for predicting and visualizing motion artifacts that are often seen when viewing digital displays. The toolbox enables users to input parameters including the stimulus configuration, the properties of the display, and viewing parameters. They can then determine whether motion artifacts will be seen and, if they are, which ones will be seen. By adjusting input parameters, the user can determine how best to minimize or eliminate the artifacts. Hopefully, this tool will aid the development of future displays. 

BiPMAP could in future be further extended to include other parameters such as retinal eccentricity, chromatic vs luminance variation, and refractive error \cite{mantiuk2022stelacsf, watson2018csf, mullen1985contrast, barten2003formula}. In addition, more complicated stimuli could be added at the front end: i.e., natural video. Finally, image-quality metrics could be included so that the user can obtain a quantitative estimate of the quality of the display being prototyped \cite{venkataramanan2021hitchhiker,nafchi2016mean,prashnani2018pieapp}. We intend to add these features in future additions of the toolbox. 

\ifCLASSOPTIONcaptionsoff
  \newpage
\fi



\bibliographystyle{IEEEtran}
\bibliography{IEEEabrv,./bare_jrnl.bib}

\end{document}